\documentclass[a4paper]{jpconf}

\pdfoutput=1

\usepackage{cmap}
\usepackage[english]{babel}
\usepackage{ucs}
\usepackage[utf8x]{inputenc}
\usepackage[T1]{fontenc}
\usepackage{lmodern}

\usepackage{microtype}

\usepackage{amsmath}
\usepackage{amssymb}
\usepackage{nicefrac}
\usepackage{mathtools}

\usepackage[dvips]{graphicx}

\usepackage[dvipsnames]{xcolor}

\usepackage[breaklinks=true,colorlinks=true,linkcolor=blue,citecolor=magenta,urlcolor=Green]{hyperref}

\usepackage[hyperpageref]{backref}
\renewcommand*{\backref}[1]{}
\renewcommand*{\backrefalt}[4]{%
   \ifcase #1 %
      { {\scriptsize Not cited in the text.}}%
   \or
      { {\scriptsize Back to page} #2}%
   \else
      { {\scriptsize Back to page} #2}%
   \fi
}

\usepackage{cite}

\begin{document}

\title{Right-handed neutrino magnetic moments}

\author{Alberto Aparici${}^1$, Kyungwook Kim${}^2$, Arcadi Santamaria${}^1$ and José Wudka${}^2$}

\address{${}^1$ Departament de Física Teòrica, Universitat de València and IFIC, Universitat de València-CSIC, 
Doctor Moliner 50, E-46100 Burjassot, València, Spain}
\address{${}^2$ Department of Physics and Astronomy, University of California, Riverside, California 92521-0413, USA}

\ead{alberto.aparici@uv.es}

\begin{abstract}
 We consider the most general dimension 5 effective Lagrangian that can be built using only Standard Model fields plus right-handed neutrinos, and find that there exists a term that provides electroweak moments (i.e., couplings to the $Z$ and photon) for the right-handed neutrinos. Such term has not been described previously in the literature. We discuss its phenomenology and the bounds that can be derived from LEP results and from the observation of the cooling process of red giants and supernovae.
\end{abstract}

\section{Motivation}

Neutrino physics has been a hot topic of research and discussion in the last thirty years, and especially since we have compelling evidence that the structure of their masses is highly nontrivial (for a review on the subject, see \cite{NakamuraPetcov:2010pdg,King:2003jb}). The remarkable smallness of these masses, at least a factor $10^5$ lighter than the electron mass, is usually regarded as an indication that new physics should be involved in their generation.

The definite nature of this new physics, of course, depends on the specific model one wishes to consider, and there are plenty of them: the seesaw mechanism \cite{King:2003jb}, which is one of the most popular proposals, the several models for radiative generation of masses \cite{Babu:1989fg} and many more. Precisely this 	abundance of proposals makes appealing the possibility of studying the new physics associated to neutrinos in a model-independent way. This can be done most easily if the new particles are very heavy, by eliminating them from our description of low-energy physics; the result is called an \emph{effective theory}\footnote{For an example of the procedure, see, for instance, \cite{Bilenky:1993bt}.}, and has been for long a powerful tool for examining neutrino physics (see, for example, two early but interesting applications in \cite{Weinberg:1979sa,Weldon:1980gi}).

However, in our current situation, and given the present knowledge and unknowns about neutrino masses, a piece of the puzzle has been mainly ignored: as we don't know if the neutrino masses are Dirac or Majorana, right-handed neutrinos might be among the light degrees of freedom of the theory. If they are, they must be included in the effective theory, or otherwise it will be incomplete. In any case, we don't know yet the nature of neutrino masses, so it seems sensible to include them for the sake of generality.

This constitutes our starting point: we want to inspect an effective theory which can describe all possible neutrino mass structures, and see what insight it can cast upon new physics effects; for that aim we include right-handed neutrinos among the low-energy fields. This we do in section \ref{sec-lagrangiano}, and promptly find a dimension five operator that describes Majorana electroweak moments for the right-handed neutrinos. This operator had not been previously discussed in the literature. We spend the remaining of this work, namely sections \ref{sec-bounds} and \ref{sec-prospects}, analysing the bounds and prospects of observation for such interactions. All along this work we follow closely that of \cite{Aparici:2009fh}, where the reader may find wider and more detailed discussions on all the topics covered here.

\section{The effective Lagrangian} \label{sec-lagrangiano}

This section is devoted to the discussion of an effective Lagrangian capable of describing the most general neutrino masses. For that purpose, we consider as low-energy fields those of the Standard Model (SM) plus a number of right-handed neutrinos. Following the rules of effective field theory, we construct with our low-energy fields all the terms compatible with our low-energy symmetries (namely, the SM symmetries). We don't have to stop at the renormalisable operators, but can continue to build operators of dimension five, six, etc.; such operators have couplings suppressed by inverse powers of a certain energy scale, the `new physics scale', that we denote by $\Lambda_\mathrm{NP}$. This scale is identified as the mass scale of the heavy particles that we have abstracted from the theory; it is the trace that these particles leave in the effective theory.

Among all the operators found, we want to focus in this work on a certain dimension five effective operator:
\begin{equation} \label{operator-flavour}
 \mathcal{L}_\zeta = \overline{\vphantom{\raisebox{0.6ex}{a}} \nu^\mathrm{c}_{\mathrm{R}}} \, \zeta \, \sigma^{\mu \nu} \nu_{\mathrm{R}} \, B_{\mu \nu} + \mathrm{H.c.}
\end{equation}
which describes Majorana electroweak moments for the right-handed neutrinos. The field $B$ in \eqref{operator-flavour} is the gauge boson of the SM hypercharge, $B_\mu = \cos \theta_W \, A_\mu - \sin \theta_W \, Z_\mu$, while $\zeta$ is a matrix of dimensionful couplings in the right-handed neutrino flavour space. Fermi statistics impose that $\zeta$ is antisymmetric, and so the electroweak moments are all \emph{transition} moments. As the operator is five-dimensional, the couplings will be suppressed by one power of the new physics scale; we will generically write $\zeta \sim \nicefrac{1}{\Lambda_{\mathrm{NP}}}$.

Before we go on with the analysis of the effective interaction, we want to present \eqref{operator-flavour} also in terms of the mass eigenfields. In the course of this work we will try to avoid assuming particular mass structures for the neutrinos; sometimes, however, the calculations will require us to do so. In those cases, we will assume a seesaw-like scenario for neutrino masses, and will use the following form for the electroweak moments:
\begin{equation} \label{operator-mass}
  \mathcal{L}_\zeta = \left( \overline{\vphantom{\raisebox{0.85ex}{a}} N} - \overline{\vphantom{\raisebox{0.25ex}{a}} \nu} \, \epsilon^* \right) \left( \zeta \, \mathrm{P_R} - \zeta^* \, \mathrm{P_L} \right) \, \sigma^{\mu \nu} \left(  N - \epsilon^\mathrm{T} \,  \nu \right) \, B_{\mu \nu}
\end{equation}
where $\nu$ represents the light, usual neutrinos, whereas $N$ represents the heavy neutrinos, which under seesaw are composed mainly of $\nu_\mathrm{R}$; the extent to which light and heavy neutrinos are mixed is measured by the matrix $\epsilon$, which is composed of small entries, and we can write generically $\epsilon \sim \sqrt{\nicefrac{m_\nu}{m_N}}$, where $m_\nu$ represents the order of the light neutrino masses, $m_\nu \sim \mathcal{O} (1 \; \mathrm{eV})$, and $m_N$ represents the order of the heavy neutrino masses. Notice that through this change of basis, the right-handed neutrino moments generate a whole set of heavy-heavy, light-light and heavy-light electroweak moments, each suppressed by the appropriate number of powers of $\epsilon$. Note also that the change of basis does not change the character of the moments: all the couplings in \eqref{operator-mass} are transition moments.

\section{Bounds on the electroweak moment} \label{sec-bounds}

In this section we review the most stringent bounds that can be placed upon the new electroweak moments, both arising from collider experiments and astrophysical observations. Several astrophysical scenarios allow us to constrain tightly the new interactions if the masses of the particles associated with the right-handed neutrinos are light. Collider bounds extend to heavier masses, but are poorer. Further comments on other bounds, either from collider or astrophysics, can be found in \cite{Aparici:2009fh}.

\subsection{Collider bounds from invisible $Z$ decays}

$Z$ boson decays were thoroughly studied at CERN's electron-positron collider LEP. In particular, comparing the width of the $Z$ line-shape with the width deduced from observed $Z$-decay events, one obtains the \emph{invisible width} of the $Z$ boson \cite{PDG:2010aa}. Within the SM only neutrinos can escape unobserved, so the invisible width is interpreted as the partial decay width to neutrinos; but were the electroweak moments to exist, the $Z$ would also decay to the massive particles composed of right-handed neutrinos (if such particles are not heavier than the $Z$ itself). Then the question is whether LEP data leave room for the new moments, especially bearing in mind that the observed invisible $Z$ width is in very good agreement with the value expected from theoretical SM calculations with three neutrino species.

To answer this question we calculate the remaining width that can not be accounted for by SM decays: from the invisible width, $\Gamma_\mathrm{inv} = 499 \pm 1.4 \; \mathrm{MeV}$, we subtract the SM contribution, which we deduce from the experimental partial width to charged leptons, $\Gamma_{\ell^+ \ell^-} = 83.984 \pm 0.086 \; \mathrm{MeV}$, a quantity that is not affected by the new moments. Using the known theoretical neutrino-to-charged-lepton width ratio, we can transform $\Gamma_{\ell^+ \ell^-}$ into a mixed theoretical-experimental quantity that we take as the SM contribution to the invisible decay width. Then:
\begin{equation} \label{delta-gamma-inv-bad}
 \Delta \Gamma_\mathrm{inv} = \Gamma_\mathrm{inv} - 3 \, \left( \frac{\Gamma_{\bar \nu \nu}}{\Gamma_{\ell^+ \ell^-}} \right)^\mathrm{SM} \, \Gamma_{\ell^+ \ell^-} = -2.6 \pm 1.5 \; \mathrm{MeV}
\end{equation}
which tells us that the contribution of new physics to the invisible width is compatible with zero. But of course a negative value for a quantity that is positive-definite is meaningless; we use the Feldman \& Cousins  prescription \cite{Feldman:1997qc} to transform \eqref{delta-gamma-inv-bad} into an upper bound on $\Delta \Gamma_\mathrm{inv}$, resulting in
\begin{equation} \label{delta-gamma-inv-good}
 \Delta \Gamma_\mathrm{inv} < 0.72 \; \mathrm{MeV} \qquad 90 \% \; \mathrm{C.L.}
\end{equation}

We can use this result to constrain our new interactions. The width of the $Z$ going to right-handed neutrinos through the electroweak moments, neglecting the masses of the products, is
\begin{displaymath}
 \Gamma \left(Z \rightarrow {\nu_\mathrm{R}}_i {\nu_\mathrm{R}}_j \right) = \frac{2}{3 \pi} \, \lvert \zeta_{ij} \rvert^2 \, \sin^2 \theta_W \, m_Z^3
\end{displaymath}
and from that, assuming that the new moments account for all of \eqref{delta-gamma-inv-good}, we have
\begin{displaymath}
 \Lambda_\mathrm{NP} = \frac{1}{\lvert \zeta_{ij} \rvert} > 7 \; \mathrm{TeV}
\end{displaymath}
This bound applies whenever $m_i + m_j < m_Z$ and gets somehow degraded if we consider massive final states with masses comparable to $m_Z$. For a picture of the excluded region, see figure \ref{figure}.

\subsection{Astrophysical bounds from the helium flash of red giants} \label{red-giants}

In 1990, Georg Raffelt pointed out \cite{Raffelt:1990pj} that the observational limits on the luminosity of red giant stars just before and after the ignition of helium could be converted into limits on the cooling mechanisms of the star core. In particular, magnetic moments for weakly interacting particles pose one of such cooling mechanisms; the reason is that photons immersed in a plasma acquire an effective mass term due to the interaction with the surrounding charged particles. Such massive photons, known as \emph{plasmons}, can decay as any other massive particle can, and if they have couplings to, say, right-handed neutrinos, those right-handed neutrinos could escape the star core, thus providing a mechanism for energy loss. The magnetic moment of any such weakly interacting particle light enough to be produced in plasmon decays is, then, bounded; specifically, it is bounded to very small values, $\mu_\mathrm{eff} < 3 \times 10^{-12} \, \mu_B$.

We have calculated the plasmon decay width to right-handed neutrinos through the electroweak moments, and	obtain, in the plasma rest frame,
\begin{displaymath}
 \Gamma \left( \gamma^* \rightarrow {\nu_\mathrm{R}}_i {\nu_\mathrm{R}}_j \right) = \frac{2}{3 \pi} \, \lvert \zeta_{ij} \rvert^2 \, \cos^2 \theta_W \, \frac{m_P^4}{\omega} = \frac{\mu_\mathrm{eff}^2}{24 \pi} \, \frac{m_P^4}{\omega}
\end{displaymath}
where $m_P$ is the plasmon effective mass, $\omega$ is the plasmon energy in the plasma rest frame, and we have neglected the mass of the final state particles. The maximum plasmon mass is given by the maximum plasma energy, which in the case of the core of red giants is around 5 keV. $\mu_\mathrm{eff}$ is the quantity bounded by \cite{Raffelt:1990pj}; from this expression we obtain that
\begin{displaymath}
 \Lambda_\mathrm{NP} > 4 \times 10^6 \; \mathrm{TeV}
\end{displaymath}
applicable whenever the mass of the final state neutrinos is lighter than the plasmon mass, that is to say, $m_i + m_j < 5 \; \mathrm{keV}$. The bound degrades slightly, due to kinematical suppression of the decay, if the masses are very close to $m_P$.

\subsection{Astrophysical bounds from the anomalous cooling of supernova cores}

The bound described in the previous section is impressively stringent, but can only apply to rather light right-handed neutrinos. As the mass of the plasmons is essentially determined by the energy of the surrounding plasma, seems wise to look for denser, hotter objects to which similar cooling limits can be applied. Among the several possibilities \cite{Raffelt:1999tx}, the one that imposes the tightest bounds is the anomalous cooling of supernova cores.

The core of a supernova is so dense that even the usual, light neutrinos become trapped; thus, a cooling mechanism for such objects requires something interacting even more weakly. Right-handed neutrinos can be a good candidate, and \eqref{operator-mass} allows us to produce them through a transition process like $\nu + \gamma \longrightarrow N$; such process has attracted much attention in the literature, and the magnetic moment involved is bounded to be $\mu_\mathrm{eff} < 3 \times 10^{-12} \, \mu_B$ \cite{Raffelt:1999tx}. The mass of the final $N$ here will be limited by the maximum energy of the plasma in the supernova core, which is around 30 MeV.

Notice that in the two previous sections we could neglect the mass of the final $\nu_\mathrm{R}$, and so forget about the mass and mixing structure of the neutrinos. The process that we are considering now, however, requires of a $\nu - N$ transition moment, which only arises from \eqref{operator-flavour} through mass mixing; therefore, to make this calculation we will have to assume something about the mass structure. As we stated in section \ref{sec-lagrangiano}, we will use the popular seesaw mechanism, which yields that the $\nu - N$ magnetic moment is to be suppressed by a factor of $\epsilon$, the heavy-light mixing, with $\epsilon \sim \sqrt{\nicefrac{m_\nu}{m_N}}$.

With all these elements, we can make the calculation and obtain the following bound:
\begin{displaymath}
 \Lambda_\mathrm{NP} > 4 \times 10^6 \times \sqrt{\frac{m_\nu}{m_N}} \; \mathrm{TeV}
\end{displaymath}
which is also depicted in figure \ref{figure}. Remember, this bound only applies for $m_N < 30 \; \mathrm{MeV}$.

\section{Prospects for the observation of the electroweak moments in colliders} \label{sec-prospects}

The electroweak moments open a new window for the production of right-handed neutrinos in colliders. These particles, being singlets under all the symmetries of the SM, are frequently regarded as `sterile'; if \eqref{operator-flavour} is present, however, the situation is different: the right-handed neutrinos can be produced through a Drell-Yan process (i.e., through an s-channel $Z$ or photon), and also can decay with very clear signatures. Of course, many searches have been conducted in the different collider experiments for such decaying neutral fermions \cite{PDG:2010aa}, all of which have yielded negative results. The bound from invisible decays of the $Z$ boson (see figure \ref{figure}) somewhat covers the excluded region provided by such searches, though for particular right-handed neutrino spectra the bounds can be extended to higher $\Lambda_\mathrm{NP}$ \cite{Aparici:2009fh}.

In any case, newer machines such as Tevatron and LHC, with enhanced capabilities both in terms of energy and luminosity, must allow us to probe further regions of the parameter space. To estimate the size of these regions, we have calculated the production cross section of heavy neutrinos in proton-proton and proton-antiproton collisions, first calculating the quark-antiquark process and then convoluting over the parton distribution functions of the proton (for which we used the CTEQ6.6M pdf set \cite{Nadolsky:2008zw}). We have translated those cross sections into explorable parameter space regions in figure \ref{figure}; in these plots, we have shaded the areas where we expect 100 or more pairs of heavy neutrinos to be produced, assuming an integrated luminosity of $10 \; \mathrm{fb}^{-1}$ for Tevatron and $1000 \; \mathrm{fb}^{-1}$ for LHC, and a center-of-mass energy of $1,96 \; \mathrm{TeV}$ for Tevatron and $14 \; \mathrm{TeV}$ for LHC.

Once produced, the heavy neutrinos must be detected; fortunately, our interactions provide a mechanism which can generate very clear signals. As we said before, all the moments in \eqref{operator-mass} are transition moments; this means that if the heavy neutrinos are to be generated in pairs, all these pairs will be composed of dissimilar particles; in particular, they will have different masses. But \eqref{operator-mass} also provides a way for massive particles to decay to lighter ones with emission of a photon; this decays are expected to be very fast, at least for $m_N \gtrsim 1 \; \mathrm{GeV}$ and $\Lambda_\mathrm{NP} < 1000 \; \mathrm{TeV}$. So, what we expect to see when a production event occurs is a cascade of hard photons with fixed, definite energies which can be easy to spot and identify.

After that, all which is left is a population of $N_1$, the lightest heavy neutrino, whose fate depends on the mass mixing with the light states; if the mixings are large, they will decay fast to SM particles and gauge bosons; if they are small, the $N_1$'s might escape the collider undetected. In between there's the possibility of the $N_1$ decaying away from the collision point, thus providing a very nice detection signal, but the analysis is not simple because the mean free path depends also on the boost of the $N_1$, which in turn depends on the kinematical conditions of the production and the energy of the parent partons. We believe that the photon cascade is a safer signal for the detection of the electroweak moments, even leaving this last possibility open.

\section{Conclusions}

We have presented here some of the results discussed in \cite{Aparici:2009fh}; we have argued that, given the current knowledge of neutrino masses, an effective field theory analysis of the Standard Model should include right-handed neutrinos, to account for possible Dirac terms in the neutrino mass matrix. By doing so, we have found a dimension five operator which provides electroweak moments for the $\nu_\mathrm{R}$; such interactions might open up a series of completely new scenarios in which the $\nu_\mathrm{R}$ are no longer `sterile', and so an analysis of their phenomenology is in order. We have discussed what we can say about these electroweak moments from past collider experiments and astrophysical observations, and presented what insight can be gained from current collider experiments.

\begin{figure}[t]
\centering
\includegraphics[width=0.8\textwidth]{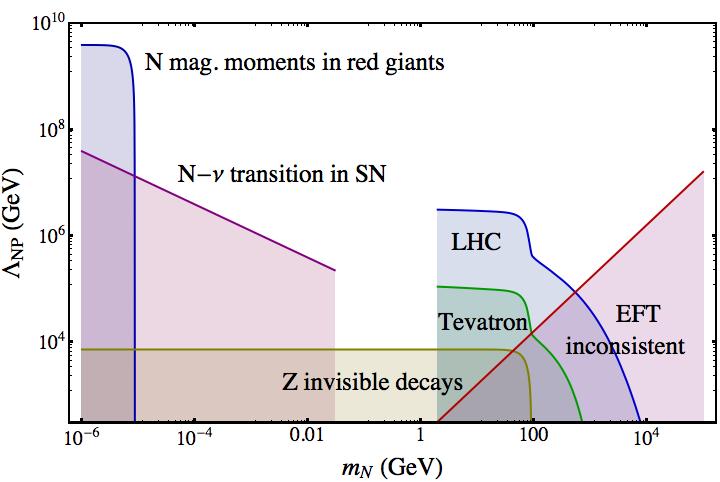} \label{figure}
 \caption{A summary of all the results presented. The shaded regions marked as `$N$ magnetic moments in red giants', `$N-\nu$ transition in SN' and `$Z$ invisible decays' represent exclusion regions, while the `Tevatron' and `LHC' ones represent regions to be probed. The rightmost red area is the region where the effective field theory loses its validity because the low-energy particles (in this case, heavy neutrinos) have masses of the order of the new physics scale.}
\end{figure}

Figure \ref{figure} summarises all these issues. We see that for right-handed neutrino masses below roughly $10 \; \mathrm{keV}$ a very stringent bound is cast upon the electroweak moments from the observational limits on energy loss in red giant cores. A milder but still tight bound can be extended up to $30 \; \mathrm{MeV}$ masses from limits on energy losses in supernova cores, but this one depends on the mixing between the right-handed and left-handed neutrinos, so it must be considered less safe and more model-dependent. Finally, the fact that the decay width of the $Z$ boson to undetected particles agrees so exactly with the theoretical expectations for the SM with three neutrino species also imposes a limit on the possible decays of $Z$ into $\nu_\mathrm{R}$; this bound is definitely weaker, but the experimental setup is solid and well understood, and it extends to masses up to $m_Z$.

Finally, the prospects for current collider experiments show that important regions of the parameter space are to be probed both in Tevatron and LHC, especially for masses below the $Z$ mass. We have explained that the collider production, via a Drell-Yan process, should be followed by a cascade of hard photons of definite energies, which constitutes a neat detection signal. Thus, we feel optimistic about the prospects for this interaction in ongoing collider experiments.

\vspace{0.4cm}

This work has been supported by the Ministry of Science and Innovation (MICINN) Spain, under Grant No. FPA2008-03373, by the European Union within the Marie Curie Research \& Training Networks, MRTN-CT-2006-035482 (FLAVIAnet), and by the U.S. Department of Energy Grant No. DE-FG03-94ER40837. A.A. is supported by the MICINN under the FPU program.

\section*{References}
\bibliography{Biblio-PASCOS}
\bibliographystyle{apsrev4.1-mod}

\end{document}